\documentclass{agnspec}
\usepackage{graphics}
\usepackage{psfig}

\def\simgt{\lower 2pt \hbox{$\, \buildrel {\scriptstyle >}\over {\scriptstyle \sim}\,$}}
\def\simlt{\lower 2pt \hbox{$\, \buildrel {\scriptstyle <}\over {\scriptstyle \sim}\,$}}

\def\chandra{{\it Chandra\/}}
\def\conx{{\it Constellation-X\/}}

\def\genx{{\it Generation-X\/}}

\def\rosat{{\it ROSAT\/}}

\def\xeus{{\it XEUS\/}}
\def\xmm{{\it XMM-Newton\/}}

\def\aox{$\alpha_{\rm ox}$}

\begin{document}

\title{Chandra and XMM-Newton Observations of the Highest Redshift 
Quasars: X-rays from the Dawn of the Modern Universe}

\authorrunning{W.N. Brandt et~al.}

\titlerunning{X-ray Observations of the Highest Redshift Quasars}

\author{W.N. Brandt,\inst{1} C. Vignali,\inst{1} X. Fan,\inst{2} S. Kaspi,\inst{3} \and D.P. Schneider\inst{1}}

\institute{Department of Astronomy \& Astrophysics, 525 Davey Laboratory, 
The Pennsylvania State University, University Park, Pennsylvania 16802, USA 
\and Institute for Advanced Study, Olden Lane, Princeton, New Jersey 08540, USA
\and School of Physics and Astronomy and the Wise Observatory, The Raymond and Beverly Sackler
Faculty of Exact Sciences, Tel-Aviv University, Tel-Aviv 69978, Israel
}

\maketitle


\begin{abstract}
We review X-ray studies of the highest redshift ($z>4$) quasars, focusing
on recent advances enabled largely by the capabilities of \chandra\ and 
\xmm. Included are discussions of basic X-ray population studies, X-ray 
spectroscopy, high-redshift AGN in X-ray surveys, and future prospects. 
\end{abstract}


\section{Introduction}

One of the main themes in astronomy over the coming decades will be the exploration
of the dawn of the modern Universe, when the first massive black holes, galaxies, and
stars formed.\footnote{See, for example, the National Research Council 2000 Decadal Report 
at http://books.nap.edu/catalog/9839.html.} X-ray 
astronomy missions such as \chandra, \xmm, \conx, \xeus, and \hbox{\genx }
can play a crucial role in this investigation by allowing studies of warm and hot 
objects in the early Universe. They will thereby complement high-redshift studies with 
observatories such as {\it ALMA\/}, {\it GSMT\/}, {\it Herschel\/}, {\it NGST\/}, and 
{\it SIRTF\/}, which will generally focus on cooler objects. For example, X-ray 
observations will permit study of the first massive black holes to form in the 
Universe; such black holes can plausibly form at $z\simgt 10$ (e.g., Haiman \& Loeb 2001). 

X-rays reveal the conditions in the immediate vicinity of the supermassive black 
hole. Measurements of the X-ray continuum's shape, amplitude relative to longer 
wavelength radiation, and variability can provide information about the inner 
accretion disk and its corona and thus ultimately offer insight into how the black 
hole is fed. The X-ray continua of quasars could plausibly change at the highest 
redshifts. For example, the rapid growth of such quasars could occur if they 
are accreting matter near the Eddington limit where accretion-disk instabilities
and ``trapping-radius'' effects (e.g., Begelman 1978) can 
arise.\footnote{Note that even models with relatively
short quasar growth times suggest that accretion rates, relative to the Eddington 
rate, are likely to increase with redshift (e.g., see Fig.~10 of Kauffmann \& Haehnelt 2000).} 
There have been some claims of X-ray continuum 
shape evolution with redshift that require further investigation 
(e.g., Vignali et~al. 1999; Blair et~al. 2000). 

The penetrating nature of X-rays allows even highly obscured black holes to be 
studied, and X-ray absorption measurements can be used to probe the environments
of high-redshift quasars. Changes in the amount of X-ray absorption
with redshift have been discussed by many authors 
(e.g., Fiore et~al. 1998; Elvis et~al. 1998; Reeves \& Turner 2000). 
The fraction of radio-loud quasars (RLQs) with heavy X-ray absorption appears to 
rise with redshift, with column densities of $\simgt 2\times 10^{22}$~cm$^{-2}$ 
being seen at $z\simgt 3$. The absorbing gas may be circumnuclear, located in the 
host galaxy, or entrained by the radio jets. The situation for the more common 
radio-quiet quasars (RQQs) is less clear due to a general lack of RQQ X-ray 
spectra at $z\simgt 2.5$. RQQs definitely show less of an absorption increase with 
redshift than do RLQs (e.g., Fiore et~al.\ 1998), but the 
constraints on an absorption/redshift connection are quite loose and require 
improvement. 

Our understanding of the X-ray emission from \hbox{$z>4$} quasars has advanced
rapidly over the past few years, enabled largely by the capabilities
of \chandra\ and \xmm. Furthermore, ground-based surveys such as the
Sloan Digital Sky Survey (SDSS; York et~al. 2000) have discovered many
new \hbox{$z>4$} quasars that are excellent \hbox{X-ray} targets. The 
number of published X-ray detections
at $z>4$ has increased to almost 30 (see Fig.~1).\footnote{See 
http://www.astro.psu.edu/users/niel/papers/highz-xray-detected.dat for
a regularly updated list of $z>4$ X-ray detections. The only $z>4$ 
non-AGN detected in the X-ray band is the \hbox{$z=4.50$} gamma-ray burst 
GRB~000131 (Andersen et~al. 2000).} This increase 
has allowed the first reliable X-ray population studies of $z>4$ quasars. 
It has also been possible to achieve X-ray detections at 
redshifts above five. Prior to 2001 the highest redshift X-ray detection 
was GB~1428+4217 at $z=4.72$ (Fabian et~al. 1997), while today there are 
nine published X-ray detections with $z=$~4.77--6.28 
(see Fig.~2; e.g., Vignali et~al. 2001; Brandt et~al. 2002). 

The rapid advances of late should continue for the next several years. 
By 2004 it is plausible that there will be $\simgt 100$ X-ray detections and 
$\simgt$~10--15 \xmm\ X-ray spectra of $z>4$ objects. 

Throughout this paper we will adopt $H_0=65$~km~s$^{-1}$ Mpc$^{-1}$, 
$\Omega_{\rm M}=1/3$, and $\Omega_{\Lambda}=2/3$. 


\begin{figure}
\centerline{\psfig{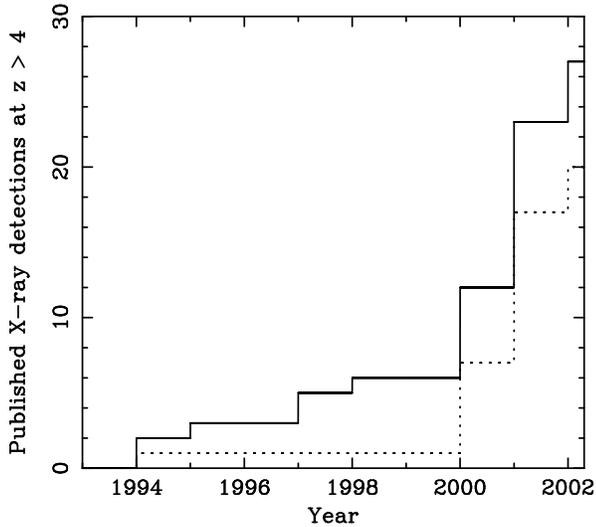}}
\caption[]{Cumulative number of published X-ray detections of $z>4$ 
quasars as a function of publication year. The solid line is for all
quasars, while the dotted line is for optically selected RQQs. 
Note the dramatic rise from 2000--2002.}
\end{figure}


\section{Current Status and Some Recent Results}

\subsection{First X-ray Population Studies at $z>4$ via Exploratory Observations}

Prior to 2000 there were only six published $z>4$ X-ray detections (see Fig.~1), 
and these were for a heterogeneous mixture of objects not suitable for population 
studies. They included three radio-loud blazars (Mathur \& Elvis 1995; 
Fabian et~al. 1997; Zickgraf et~al. 1997), two X-ray selected RQQs 
(Henry et~al. 1994; Schneider et~al. 1998), and only one optically 
selected RQQ (Bechtold et~al. 1994). The blazars were notable for
their large X-ray fluxes that allowed moderate-quality X-ray 
spectroscopy (see Fig.~3; Moran \& Helfand 1997; Fabian et~al. 1998;
Boller et~al. 2000; Yuan et~al. 2000; Fabian et~al. 2001a; 
also see Fabian et~al. 2001b for the X-ray spectrum of a 
recently discovered $z>4$ blazar). All of the four $z>4$ blazars
currently detected in the X-ray regime appear to have their  
emission dominated by jets. Three of the four show evidence for
low-energy spectral flattening that may be due to intrinsic absorption, 
in line with the findings for lower redshift RLQs described in \S1. 
While these blazars have been wonderful targets for study, they
are not representative of the majority quasar population at $z>4$
(e.g., Stern et~al. 2000). Furthermore, due to the strength of 
the X-ray emission from the jet, it is difficult to observe the 
accretion disk and the immediate vicinity of the black hole. 


\begin{figure}
\centerline{\psfig{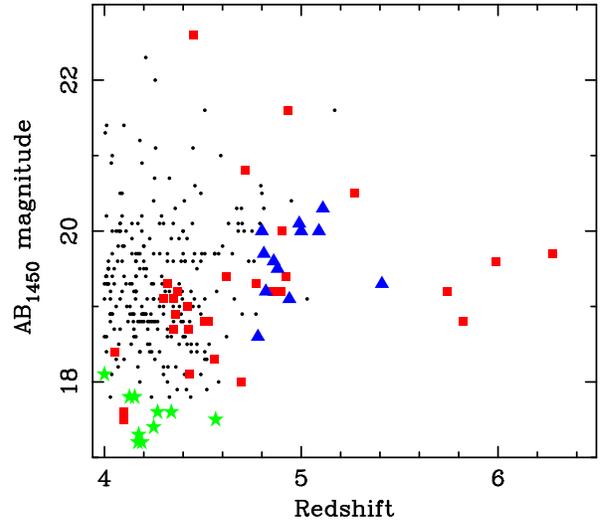}}
\caption[]{$AB_{1450}$ magnitude versus redshift for $z>4$ quasars
(small dots). Solid squares indicate quasars that have published 
X-ray detections or tight upper limits at present. Triangles and 
stars indicate SDSS and PSS quasars, respectively, that we are 
observing during \chandra\ Cycle~3.}
\end{figure}


With the 2000--2002 studies, optically selected RQQs, the majority 
quasar population, now dominate the number of $z>4$ X-ray 
detections (see Fig.~1). There are 20 such detections published at 
present. Six of these were obtained in an archival study of \rosat\ data 
by Kaspi, Brandt, \& Schneider (2000), and an additional 12 were
obtained via exploratory (2--10~ks) \chandra\ observations 
(e.g., Vignali et~al. 2001; Brandt et~al. 2002; 
Mathur, Wilkes, \& Ghosh 2002; Schwartz 2002b).\footnote{The 
other two optically selected RQQs with published X-ray 
detections are Q0000--2619 (Bechtold et~al. 1994; detected by \rosat) and 
SDSS~1044--0125 (Brandt et~al. 2001a; detected by \xmm).} We have ongoing 
\chandra\ programs designed to obtain additional detections;
\chandra\ is extremely effective at this and, near the
Advanced CCD Imaging Spectrometer (ACIS; G.P. Garmire et~al., in 
preparation) aim point, can detect 
quasars with as few as 2--3 counts! As shown 
in Fig.~2, we are focusing on $z>4.8$ quasars from the SDSS
and optically bright quasars from the Palomar Digital Sky Survey 
(PSS; e.g., Djorgovski et~al. 1998).\footnote{See 
http://www.astro.caltech.edu/$\sim$george/z4.qsos for a listing of the 
PSS quasars.} We have obtained a significant number of new 
detections, and these will be published shortly (C. Vignali et~al., 
in preparation). 

The exploratory observations have defined the typical X-ray fluxes
and luminosities of $z>4$ quasars. As shown in Fig.~3, these quasars
are typically faint X-ray sources, and even the brightest non-blazars
generally have \hbox{0.5--2~keV} fluxes $\simlt 3\times 10^{-14}$~erg~cm$^{-2}$~s$^{-1}$. 
The practical implication of this is that X-ray spectroscopy of $z>4$ quasars
will generally be a challenging endeavor (see \S2.2 for details). 
There is a clear correlation between X-ray flux and $AB_{1450}$ magnitude,
although there is significant scatter around this correlation (see Fig.~3;
Vignali et~al. 2001 give the results of correlation tests). The scatter
makes it risky to attempt long spectroscopic observations before an 
X-ray flux has been established via exploratory observations. 
The 2--10~keV rest-frame luminosities of the observed quasars range from
$\approx$~$2\times 10^{44}$~erg~s$^{-1}$ to 
$\approx 4\times 10^{45}$~erg~s$^{-1}$. 


\begin{figure}
\centerline{\psfig{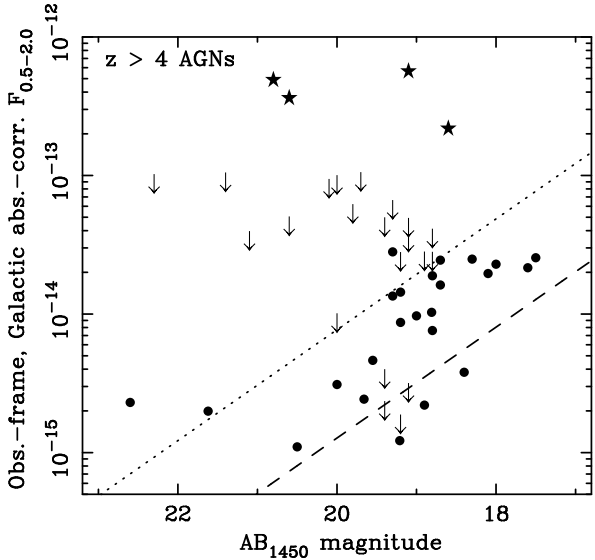}}
\caption[]{Observed-frame, Galactic absorption-corrected \hbox{0.5--2~keV} flux 
versus $AB_{1450}$ magnitude for $z>4$ AGNs; the units of the ordinate are 
erg~cm$^{-2}$~s$^{-1}$. The dots and arrows show $z>4$ detections 
and upper limits, respectively, from Kaspi et~al. (2000), 
Brandt et~al. (2001a), Vignali et~al. (2001), Silverman et~al. (2002), and 
Brandt et~al. (2002). Blazars at $z>4$ are shown as stars. The slanted lines 
show $z=5.0$ loci for $\alpha_{\rm ox}=-1.5$ (dotted) and 
$\alpha_{\rm ox}=-1.8$ (dashed).}
\end{figure}


The exploratory observations have also allowed a basic assessment of the
contribution of X-ray emission to the broad-band spectral energy distributions
(SEDs) of $z>4$ quasars. Fig.~4, for example, shows the quantity \aox\ plotted
versus redshift for optically selected RQQs; \aox\ is the slope of a nominal 
power law between 2500~\AA\ and 2~keV in the rest frame 
[$\alpha_{\rm ox}=0.384\log (f_{\rm 2~keV}/f_{2500~\mbox{\footnotesize \AA}}$) 
where $f_{\rm 2~keV}$ is the flux density at 2~keV and
$f_{2500~\mbox{\footnotesize \AA}}$ is the flux 
density at 2500~\AA].\footnote{For the $z>4$ quasars in Fig.~4 we have 
generally calculated $f_{\rm 2~keV}$ from the observed-frame 0.5--2~keV
flux (assuming an \hbox{X-ray} power-law photon index of $\Gamma=2$); this is
required because ACIS has limited spectral response below 0.5~keV. 
Thus, the derived \aox\ values are actually based on the relative
amount of X-ray flux in the 0.5$(1+z)$~keV to 2$(1+z)$~keV rest-frame band.}
There is no evidence for strong changes in \aox\ with redshift, despite 
the large changes in quasar number density over the redshift range shown
in Fig.~4 (see \S5.3 of Fan et~al. 2001 and references therein).
Some quasars at $z>4$ may have slightly more negative \aox\ values, perhaps due 
to X-ray absorption by the large amounts of gas in their primeval 
host galaxies (see Vignali et~al. 2001 for further discussion). 
However, the \aox\ constraints at $z>4$ (and especially at $z\simgt 4.8$) 
require improvement via further observations before small \aox\ changes 
can be considered significant. 
The lack of strong changes in \aox\ with redshift is generally consistent 
with the lack of strong spectral evolution at other wavelengths. For 
example, the ultraviolet rest-frame spectra of $z>4$ quasars are only
subtly different from those at lower redshifts 
(e.g., Schneider, Schmidt, \& Gunn 1989; Constantin et~al. 2002), 
and the fraction of RLQs does not appear to change significantly with 
redshift (e.g., Stern et~al. 2000). 


\begin{figure*}
\vspace*{-0.7in}
\centerline{\psfig{figure=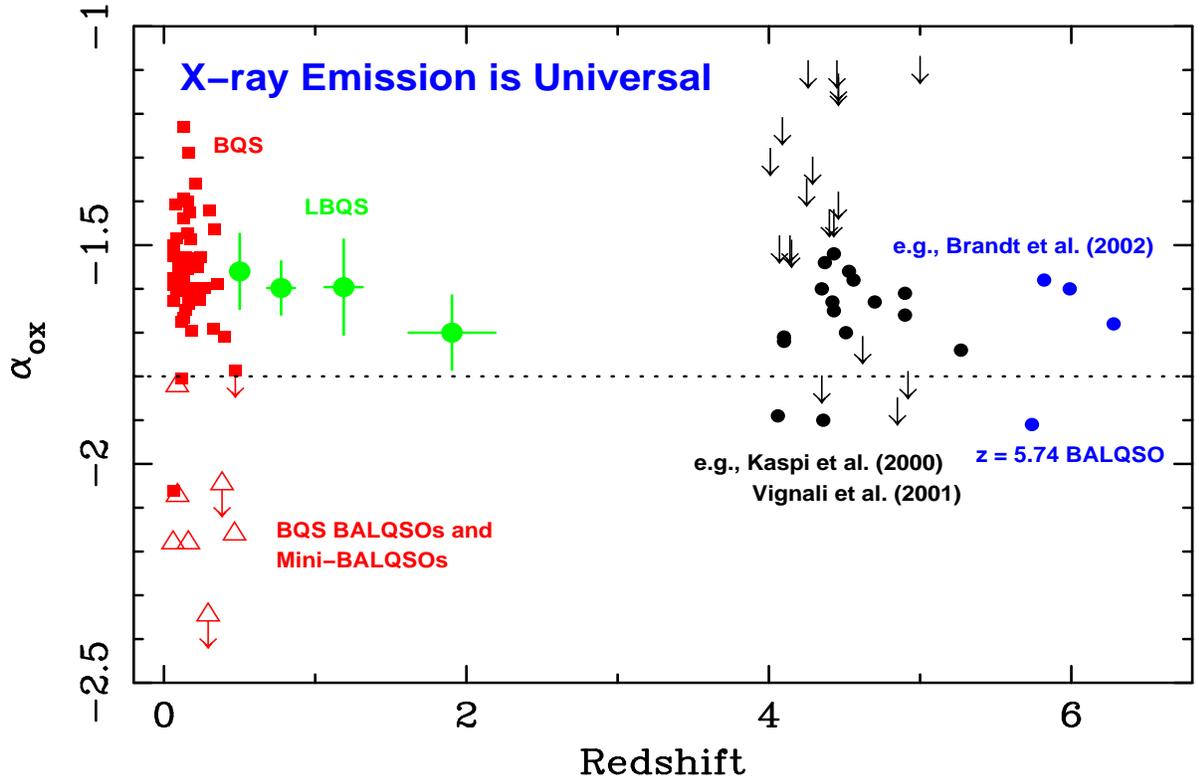,height=14.0cm,width=19.4cm,angle=-90}}
\vspace*{-0.7in}
\caption[]{The parameter \aox\ versus redshift 
for optically selected RQQs. The open triangles are for
seven luminous, absorbed Bright Quasar Survey (BQS; Schmidt \& Green 1983) 
RQQs, and the solid squares are for the other 46 luminous BQS RQQs (from 
Brandt et~al. 2000). The large solid dots with error bars show stacking 
results for Large Bright Quasar Survey (LBQS; Hewett, Foltz, \& Chaffee 1995) 
RQQs from Figure~6d of Green et~al. (1995). The small solid dots and plain arrows
show $z>4$ detections and upper limits, respectively, from Kaspi et~al. (2000), 
Brandt et~al. (2001a), Vignali et~al. (2001), and Brandt et~al. (2002). A 
horizontal line has been drawn at $\alpha_{\rm ox}=-1.8$ to guide the eye.}
\end{figure*}


Fig.~4 has several important implications. 
First of all, it supports the idea that X-ray emission is a universal 
property of quasars at all redshifts; this result had already been 
established for low-redshift quasars ($z\simlt 0.5$; 
e.g., Avni \& Tananbaum 1986; Brandt, Laor, \& Wills 2000). Most of 
the X-ray nondetections in Fig.~4 with $\alpha_{\rm ox}<-1.8$ are 
Broad Absorption Line (BAL) QSOs or mini-BALQSOs; there is now 
evidence that these quasars indeed produce X-rays at a typical level 
but that this emission is absorbed (e.g., Gallagher et~al. 2001; 
Green et~al. 2001; Gallagher et~al. 2002). Thus, Fig.~4 bodes well 
for attempts to detect the first massive black holes to form in 
the Universe ($z\approx$~8--20) with deep X-ray surveys; the current  
data suggest that these objects are likely to be luminous X-ray emitters. 
Secondly, the fact that \aox\ does not change strongly with redshift helps 
to validate the bolometric correction factor usually adopted 
(e.g., Fan et~al. 2001) when estimating the black hole masses of 
high-redshift quasars via the Eddington argument.
The derived masses of $\approx$~(1--5)$\times 10^9$~M$_\odot$ place 
constraints on theories of cosmic structure formation 
(e.g., Efstathiou \& Rees 1988; Turner 1991).
Thirdly, the X-ray emission of high-redshift quasars is expected to 
affect the heating and ionization of the intergalactic medium 
(e.g., Venkatesan, Giroux, \& Shull 2001).\footnote{Due to 
its low metallicity and relatively low integrated column 
density, the intergalactic medium should not prevent the 
X-ray detection of quasars even to very high redshift 
(e.g., Aldcroft et~al. 1994; Weinberg et~al. 1997; Miralda-Escud\'e 2000).} 
The data in Fig.~4 suggest that, at least to first order, the SEDs of local 
quasars may be adopted when trying to compute the effects of early quasars
upon the intergalactic medium. 

Some studies have found evidence that \aox\ depends upon quasar optical 
luminosity, with more luminous quasars generally having more negative
values of \aox\ (e.g., Avni, Worrall, \& Morgan 1995; Green et~al. 1995; 
but see Yuan, Siebert, \& Brinkmann 1998 for an alternative point of 
view). We have searched for such a trend using the RQQs shown in Fig.~4;
BALQSOs and mini-BALQSOs have been removed since their \aox\ values
are affected by absorption. A weak trend is possible, but additional 
observations (of a well-defined RQQ sample) will be required to determine 
if one is present since there is significant scatter in \aox\ 
($\approx\pm 0.2$) at all optical luminosities (see \S4.2 of 
Vignali et~al. 2001 for details). It is 
worth noting that \hbox{$\approx$~5--30\%} of
$z\approx$~4--6 quasars are expected to suffer from strong flux
amplification (by a factor of $\simgt 2$) due to gravitational 
lensing (e.g., Wyithe \& Loeb 2002). Such an effect will confuse 
studies of the dependence of \aox\ upon luminosity; \aox\ will 
probably not be changed by gravitational lensing, but the true 
luminosity will differ from that assumed. 

Finally, exploratory observations of $z>4$ quasars allow searches
for extended X-ray emission associated with jets. Schwartz (2002a) has 
argued that X-ray jets can serve as cosmic beacons. If the 
X-ray emission from jets is produced via the Compton scattering of 
cosmic microwave background (CMB) photons by relativistic electrons, 
then it should be detectable to high redshift; the increase in the 
CMB energy density with redshift compensates for cosmological 
surface-brightness dimming. Schwartz (2002b) has recently 
identified a possible jet from the quasar SDSS~1306+0356;
notably, this quasar is not a strong radio source. Further
observations are required to check this possibility, and exploratory
observations of a large number of $z>4$ quasars should allow a search
for an overdensity of ``companion'' X-ray sources that could be 
radio jets. 

\subsection{X-ray Spectral Analyses at $z>4$}

Aside from the four blazars mentioned near the start of \S2.1, detailed
X-ray spectral analyses of $z>4$ quasars have not been performed at
present. Such analyses are important to search for changes in the
X-ray power-law photon index and X-ray absorption 
(neutral or ionized) with redshift. To our knowledge, the 
highest redshift RQQ with X-ray spectral constraints of reasonable 
quality is the $z=3.87$ gravitationally lensed BALQSO APM~08279+5255 
(see Fig.~5; Gallagher et~al. 2002). This object has a power-law 
photon index of $\Gamma=1.86^{+0.26}_{-0.23}$ (for the rest-frame
5--25~keV band), consistent with those seen for lower-redshift 
RQQs (e.g., George et~al. 2000; Reeves \& Turner 2000). 

The lack of quasar X-ray spectroscopy at $z>4$ is largely due to 
their low X-ray fluxes which limit the number of photons that
can be gathered (see \S2.1). Basic hardness ratios for several 
objects appear consistent with those of low-redshift quasars
(e.g., Vignali et~al. 2001), but even these have large errors. 
With \hbox{40--100~ks} exposures, \xmm\ 
can obtain moderate-quality ($\approx$~1000--2000 count) 
$\approx$~1--50~keV spectra of $z>4$ quasars with 0.5--2~keV fluxes 
of $\simgt 1.5\times 10^{-14}$~erg~cm$^{-2}$~s$^{-1}$ 
(see Fig.~3).\footnote{These data can also be used to search for
X-ray variability. At present, highly significant X-ray variability 
at $z>4$ has only been detected from the blazars
PMN~J0525--3343 (Fabian et~al. 2001b), 
GB~1428+4217 (Fabian et~al. 1999; Boller et~al. 2000), and 
GB~1508+5714 (Moran \& Helfand 1997). 
There are tentative claims that quasar X-ray variability increases 
with redshift (e.g., Manners, Almaini, \& Lawrence 2002).}
Furthermore, stacking the counts from quasars observed by \xmm\ (as 
well as those with exploratory \chandra\ observations) can provide 
tight average spectral constraints and allow searches for spectral
features such as iron~K$\alpha$ lines.  
Effective X-ray spectroscopy of quasars with 0.5--2~keV fluxes
of $\simlt 1.5\times 10^{-14}$~erg~cm$^{-2}$~s$^{-1}$ will require 
future missions such as \conx, \xeus, and \genx; populating Fig.~3 with 
further detections will allow optimum planning of these missions. 


\begin{figure}
\centerline{\psfig{figure=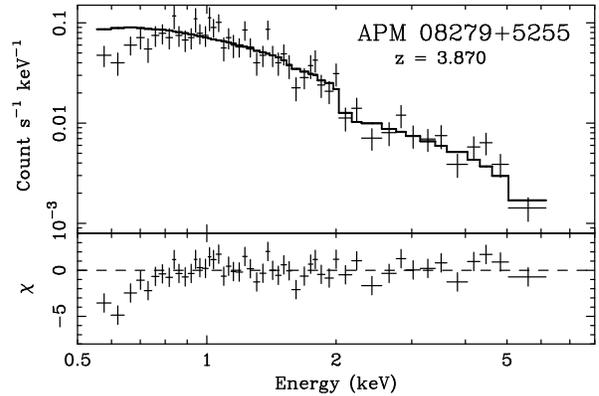,width=7.8cm,angle=-90}}
\caption[]{Observed-frame \chandra\ ACIS spectrum of the \hbox{$z=3.87$} gravitationally 
lensed BALQSO APM~08279+5255. The spectrum has been fit above rest-frame 5~keV
with a power-law model that has then been extrapolated back to lower energies. 
The lower panel shows fit residuals in units of $\sigma$ with error bars of
size one. The systematic residuals below observed-frame 1~keV are thought
to be due to intrinsic absorption ($N_{\rm H}\approx 6\times 10^{22}$~cm$^{-2}$)
associated with the BAL wind. Adapted from Gallagher et~al. (2002).}
\end{figure}


\subsection{AGN at $z>4$ in X-ray Surveys}

At present, there are four published $z>4$ quasars that were selected 
based upon their detection in X-ray surveys: 
RX~J1759+6638 (Henry et~al. 1994), 
RX~J1028--0844 (Zickgraf et~al. 1997), 
RX~J1052+5719 (Schneider et~al. 1998), and 
CXOMP~J213945.0--234655 (Silverman et~al. 2002). 
The Lockman Hole source RX~J1052+5719 is notable due to its 
low luminosity (it is the point furthest to the left in 
Fig.~3), and CXOMP~J213945.0--234655 is the highest 
redshift X-ray selected quasar published to date ($z=4.93$). 
Follow-up studies of sources from the \chandra\ Deep Field-North 
survey (Brandt et~al. 2001c) have recently discovered a still 
higher redshift quasar at $z=5.18$ 
(A.J. Barger et~al., in preparation).\footnote{See  
http://www.astro.wisc.edu/$\sim$barger/hizq.jpg for a plot
of the spectrum.} This object's spectrum is notable for 
its relatively narrow Lyman~$\alpha$ line. 

The radio-selected AGN VLA~J1236+6213 has also been detected 
in the \chandra\ Deep Field-North survey (Brandt et~al. 2001b); 
this object has a likely redshift of $z=4.424$ (Waddington et~al. 1999). 
VLA~J1236+6213 is a very faint source in both the optical and 
X-ray bands with $AB_{1450}\approx 25.2$ and a 0.5--2~keV flux 
of $\approx 9\times 10^{-17}$~erg~cm$^{-2}$~s$^{-1}$ (compare 
with Fig.~3). Its rest-frame 2--10~keV luminosity is 
$\approx 2\times 10^{43}$~erg~s$^{-1}$, comparable to those 
of the local Seyfert galaxies NGC~3516, NGC~3783, and NGC~5548. 
This is by far the lowest luminosity X-ray source known at $z>4$, 
and its detection demonstrates that \chandra\ is achieving the 
sensitivity needed to study Seyfert-luminosity AGN at high 
redshift.

A large population of Seyfert-luminosity AGN at $z\approx$~4--10 
has been postulated by Haiman \& Loeb (1999). These ``proto-quasars''
would represent the first massive black holes to form in the Universe, 
and they would play an important role in galaxy formation. 
One of NASA's long-term goals for X-ray astronomy is to understand
such objects (e.g., White 2002), and ultradeep \chandra\ exposures 
are one of the few ways that they might be found at present. With a
5--10~Ms exposure, \chandra\ can achieve sensitivities comparable
to those discussed for missions such as \xeus\ (see Fig.~6). 
A 5~Ms \chandra\ observation could detect a $z=10$ proto-quasar
down to a rest-frame 5.5--22~keV luminosity of 
$\approx 8\times 10^{42}$~erg~s$^{-1}$ (after a plausible bolometric
correction, this is the X-ray luminosity expected from an 
$\approx 10^6$~M$_\odot$ black hole radiating near the Eddington limit). 
At $z=10$ \chandra\ provides rest-frame sensitivity up to 
$\approx 90$~keV, and such high-energy X-rays can penetrate a 
substantial amount of obscuration. \chandra\ positions 
are likely to be the best available 
for at least 15--20 years, and precise positions will
be essential for reliable identification of 
proto-quasars.\footnote{\genx\ is planned to have an angular
resolution comparable to or better than that of \chandra\ (see Fig.~6). 
However, it is not expected to launch until $\approx 2020$, and 
its construction will require challenging improvements in lightweight 
precision X-ray optics. See the \genx\ white paper by Zhang et~al. 
at http://universe.gsfc.nasa.gov/docs/roadmap/submissions.html.}


\begin{figure}
\centerline{\psfig{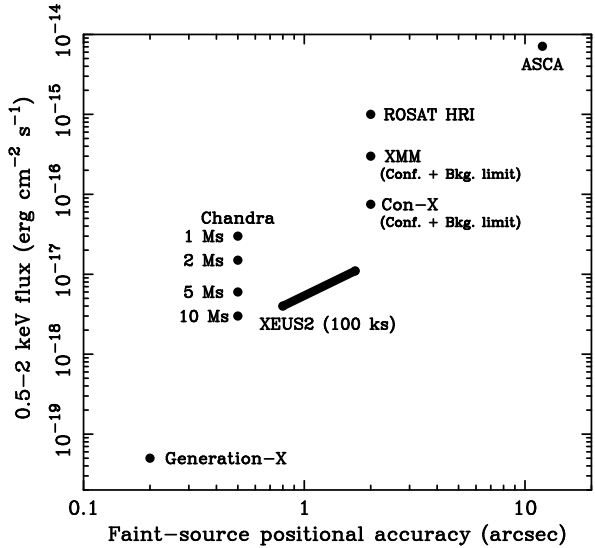}}
\caption[]{Soft-band flux limit versus faint-source positional accuracy for 
some past, present, and future X-ray missions. Note that the locations in 
the diagram for future missions should be taken as approximate, and
that \conx\ is focused on high-throughput spectroscopy rather than
deep surveys. Both \xmm\ and \conx\ are background limited and suffer
from source confusion at approximately the positions shown. With 
sufficient exposure, \chandra\ can achieve sensitivities 
comparable to those discussed for future missions 
such as \xeus. Furthermore, \chandra\ positions are likely to be 
the best available for at least 15--20~years.}
\end{figure}


At $z\simgt 6.5$ an AGN should appear optically blank; the 
Lyman~$\alpha$ forest and Gunn-Peterson trough will absorb 
essentially all of the flux through the $I$ band. Thus, an upper 
limit on the space density of proto-quasars can be set simply 
by counting the number of X-ray sources that lack any optical 
counterpart.\footnote{Of course, this method requires the 
assumption that proto-quasars be X-ray luminous. The best 
available data suggest that this should be the case (see \S2.1).} 
Effective application of this method, however, requires exceptionally 
deep optical imaging to prevent confusion between truly optically blank 
sources at extreme redshift and very optically faint sources at 
moderate redshift (e.g., objects like the $I=25.8$, $z\approx 2.75$ 
AGN CXOHDFN~J123651.7+621221; Brandt et~al. 2001b and references therein). 
At present, the \chandra\ Deep Field-North data suggest that there is 
$\simlt 1$ AGN detected at $z\simgt 6.5$ per \hbox{$\approx 12$~arcmin$^{2}$} 
(Alexander et~al. 2001; down to a \hbox{3.75--15~keV} luminosity limit 
at $z=6.5$ of $\approx 2\times 10^{43}$~erg~s$^{-1}$). This limit 
should soon be tightened substantially via improved ground-based 
imaging and the Great Observatories Origins Deep Survey 
(GOODS).\footnote{See
http://www.stsci.edu/science/goods/ for a description of GOODS.} 

A 5--10~Ms \chandra\ exposure should also allow normal and 
starburst galaxies at $z>4$ to be studied via count stacking
analyses. Such analyses using the 1~Ms data for the \chandra\ 
Deep Field-North have recently been used to determine the 
basic X-ray properties of $z=$~2--4 Lyman break galaxies 
(Brandt et~al. 2001d). 


\section{Some Future Prospects}

Further \chandra\ and \xmm\ observations can significantly improve our
understanding of $z>4$ quasars, thereby laying the groundwork for 
future high-redshift \hbox{X-ray} efforts. Some important studies include 
the following: 

\begin{enumerate}

\item
Exploratory \chandra\ observations of additional $z\simgt 4.8$ quasars 
will allow a better search for any dependence of \aox\ upon redshift 
(see Fig.~4) or luminosity. The SDSS is expected to find $\approx$~50--100 
quasars with $z>5$ and $\approx 10$ quasars with $z>6$; it can find quasars 
up to $z\approx 6.5$ (e.g., Fan et~al. 2001). Additional quasar surveys
are being implemented to search for quasars up to $z\approx 7.2$
(e.g., Warren \& Hewett 2002). 

\item
Exploratory \chandra\ observations of the optically brightest $z>4$ 
quasars known (see Fig.~2) can identify further targets appropriate for
\xmm\ spectroscopy. 

\item
The SDSS has identified some unusual quasars lacking emission lines
(e.g., Fan et~al. 1999; Anderson et~al. 2001). The nature of these
objects is mysterious, and defining their basic X-ray properties may
provide insight. The one object observed thus far \hbox{(SDSS~1532--0039)}
was not detected and appears to be fairly X-ray weak (Vignali et~al. 2001). 

\item
The X-ray properties of $z>4$ BALQSOs are poorly understood. There
has been speculation that the \hbox{X-ray} absorption in BALQSOs may increase
with redshift (e.g., \S4.3 of Vignali et~al. 2001), but better \aox\
constraints and X-ray observations of more BALQSOs are required. A large
negative value of \aox\ can help to identify a BALQSO when BALs cannot
be identified due to limited rest-frame ultraviolet spectral coverage
(e.g., Brandt et~al. 2001a; Goodrich et~al. 2001; Maiolino et~al. 2001). 

\item
The X-ray properties of $z>4$ RLQs are poorly defined. 
There are few X-ray observations of $z>4$ quasars that are intermediate
in radio loudness between RQQs and highly radio-loud blazars (see Fig.~3). 
Filling this radio-loudness gap is important, and some of these radio-loud
objects may be bright enough for \xmm\ spectroscopy. 

\item
\xmm\ can obtain moderate-quality ($\approx$~1000--2000 count) 
$\approx$~1--50~keV spectra of $z>4$ quasars with 0.5--2~keV fluxes 
of $\simgt 1.5\times 10^{-14}$~erg~cm$^{-2}$~s$^{-1}$ (see Fig.~3). 
These observations will constrain changes in the X-ray power-law photon 
index and X-ray absorption (neutral or ionized) with redshift. Furthermore, 
stacking the counts from quasars observed by \xmm\ (as well as those with 
exploratory \chandra\ observations) can provide tight average spectral 
constraints and allow searches for spectral features such as 
iron~K$\alpha$ lines.  

\item 
A 5--10~Ms \chandra\ survey can effectively search for the first massive
black holes to form in the Universe (see Fig.~6). 

\end{enumerate}

\noindent
In the more distant future, \conx, \xeus, and \genx\ should allow 
detailed X-ray studies of the highest redshift quasars
(see \S5 of Vignali et~al. 2001 for some simulations). 


\begin{acknowledgements}
We thank all of our collaborators on the work reviewed here. 
We thank D.M. Alexander, F.E. Bauer, S.C. Gallagher, A.E. Hornschemeier, 
and M.A. Strauss for helpful discussions. 
We gratefully acknowledge the financial support of 
NASA LTSA grant NAG5-8107 (WNB, CV, SK),
\chandra\ X-ray Center grant G01-2100X (WNB, CV, SK),
NSF grant PHY00-70928 (XF), and 
NSF grant AST99-00703 (DPS). 
\end{acknowledgements}



\end{document}